\documentclass[fleqn,usenatbib]{mnras}

\usepackage{newtxtext,newtxmath}

\usepackage[T1]{fontenc}

\DeclareRobustCommand{\VAN}[3]{#2}
\let\VANthebibliography\thebibliography
\def\thebibliography{\DeclareRobustCommand{\VAN}[3]{##3}\VANthebibliography}

\usepackage{graphicx}	
\usepackage{amsmath}	
\usepackage{amssymb}	
\usepackage{media9}

\newcommand{\hnii}{{\rm H}$\alpha+$[N~{\sc ii}]}
\newcommand{\oiii}{[O\,{\sc iii}]}

\newcommand{\HNII}{{\rm H}$\alpha+$[N~{\sc ii}]~6548, 6584~\AA}

\newcommand{\OIII}{[O {\sc iii}]~5007~\AA}
\newcommand{\SII}{[S~{\sc ii}]~6716, 6731~\AA}
\usepackage{soul}

\title[First 3D MK model of SNRs: VRO 42.05.01]{First 3D Morpho-Kinematic model of Supernova Remnants.\\
The case of VRO 42.05.01 (G 166.0{\it +}4.3)}

\author[Derlopa et al.]{S. Derlopa,$^{1,2}$\thanks{E-mail: sophia.derlopa$@$noa.gr}
P. Boumis,$^{1}$\thanks{E-mail: ptb$@$astro.noa.gr} 
A. Chiotellis,$^{1}$
W. Steffen,$^{3}$
S. Akras,$^{4}$
\\
$^{1}$Institute for Astronomy, Astrophysics, Space Applications
and Remote Sensing, National Observatory of Athens,
GR 15236 Penteli, Greece\\
$^{2}$Department of Physics, National and Kapodistrian University of Athens, Panepistimiopolis, GR 15783 Zografos, Greece\\
$^{3}$Instituto de Astronom\'ia, Universidad Nacional Aut\'onoma de M\'exico, Ensenada 22800, Baja California, Mexico\\
$^{4}$Instituto de Matem\'{a}tica, Estat\'{i}stica e F\'{i}sica, Universidade Federal do Rio Grande, Av. Italia km 8, 96203-900, Rio Grande, Brazil\\}

\date{Accepted 2020 August 03. Received 2020 August 02; in original form 2020 June 17}

\pubyear{2020}

\begin{document}
\label{firstpage}
\pagerange{\pageref{firstpage}--\pageref{lastpage}}
\maketitle

\begin{abstract}
We present the first three dimensional (3D) Morpho-Kinematic (MK) model of a supernova remnant (SNR), using as a case study the Galactic SNR VRO 42.05.01. We employed the astrophysical code SHAPE in which wide field imaging and high resolution spectroscopic data were utilized, to reconstruct its 3D morphology and kinematics. We found that the remnant consists of three basic distinctive components that we call: a \lq\lq shell\rq\rq, a \lq\lq wing\rq\rq ~and a \lq\lq hat\rq\rq. With respect to their kinematical behaviour, we found that the \lq\lq wing\rq\rq ~and the \lq\lq shell\rq\rq\ ~have similar expansion velocities ($V_{\rm exp}$~=~115$\pm 5~$\,km\,s$^{-1}$). The \lq\lq hat\rq\rq\ ~presents the lowest expansion velocity of the remnant ($V_{\rm exp}$~=~90$\pm 20~$\,km\,s$^{-1}$), while the upper part of the \lq\lq shell\rq\rq\ ~presents the highest velocity with respect to the rest of the remnant ($V_{\rm exp}$~=~155$\pm 15~$\,km\,s$^{-1}$). Furthermore, the whole nebula has an inclination of $\sim$3$\degr$ - $5$\degr\ with respect to the plane of the sky and a  systemic velocity of $V_{\rm sys}=$ -17$\pm 3~$\,km\,s$^{-1}$. We discuss the interpretation of our model results regarding the origin and evolution of the SNR and we suggest that VRO 42.05.01 had an interaction history with an inhomogeneous ambient medium most likely shaped by the mass outflows of its progenitor star.
\end{abstract}

\begin{keywords}
Supernova Remnants: general -- Individual objects: VRO 42.05.01 (G 166.0$+$4.3) -- ISM: kinematicals and dynamics, 3D modelling
\end{keywords}



\section{Introduction}
\label{Introduction}

Massive stars (M $\geq$ 8 M$_{\sun}$) and carbon oxygen white dwarfs members of interacting binaries (i.e. \citealt{FIL1997}) may undergo an explosively violent death (supernova explosion, SN). The SN ejecta expand and sweep up the ambient medium.  The resulting structure is progressively transformed into a beautiful gaseous nebula which is called supernova remnant (SNR). SNRs chemically enrich the host galaxy, they influence its dynamics, while the shock waves generated after the explosion are efficient cosmic ray accelerators. Moreover, the large-scale of asymmetries and complex structures that SNRs usually display, reveal inhomogeneities present in the ambient medium where they evolve in, and provide clues about the progenitor star, since SNRs interact with the material expelled during the progenitor's evolution (\citealt{McKee1988, Chiotellis12}). The importance and utility of probing SNRs relies on providing answers for the above crucial astrophysical topics.

Valuable information about the physical processes that dominate in SNRs are gained through imaging and spectroscopic data. Nevertheless, regardless of the details that state-of-the-art astronomical instruments can depict, the fact that these data are two dimensional (2D) restricts the range of our knowledge for these objects, due to the absence of the information in the third dimension along the light of sight. Consequently, the benefits from a three dimensional (3D) study of a SNR aim at a deeper interpretation of the collected observational data.

Up to date, there are two main important tools for gaining information on the missing third dimension of SNRs: (a) the 3D (magneto) hydrodynamic (MHD) models (\citealt{TOL2014G352}; \citealt{BOLT2015}; \citealt{ABE2017}; \citealt{POT2014}; \citealt{ORL2019}), which reproduce the 3D physical properties by comparing the 2D projection of the models with the observational data, and (b) the 3D velocity-maps (\citealt{DEL2010}; \citealt{ALA2014}; \citealt{MIL2013}; \citealt{WIL2017}) which are created by the proper motion and Doppler shifted velocities of different parts of the remnant.

In this paper we take a third approach, the so-called "Morpho-Kinematic modeling" which, up to now, it has been applied successfully in Planetary Nebulae (\citealt{akras2012a}; \citealt{akras2012b}; \citealt{CLY2015}; \citealt{akras2016}; \citealt{FANG2018}; \citealt{DER2019}; \citealt{Gordillo2020}), but never in SNRs. This method reconstructs the 3D morphology of the object by using imaging and high-resolution spectroscopic data. The lack of such a 3D model in the field of SNRs was our motivation to proceed in the creation of the first 3D Morpho-Kinematical (MK) model of a SNR. For our 3D MK model the astrophysical software SHAPE was employed \citep{STEFF2017}, while as a case study we used the Galactic SNR VRO 42.05.01 (hereafter VRO) for which no 3D model has been constructed before.

The paper is organised as follows. The VRO properties are presented in Section 2. The observations and data analysis are described in Section 3. The 3D MK modelling and its results are presented in Section 4 and 5, respectively. In Section 6 we discuss the interpretation of our results and we end with our conclusions in Section 7.

\section{SNR VRO 42.05.01}
\label{sec:SNR VRO 42.05.01}
VRO 42.05.01 (G 166.0$+$4.3, \citealt{DICK1965}) is a well studied Galactic, mixed-morphology SNR (\citealt{BOUMIS2016}; 2020 in prep; \citealt{ARI2019a, ARI2019b} and references therein). This remnant was chosen for the 3D MK model due to its intriguing morphology which basically consists of two main parts: i) a hemisphere at the northeastern region called the \lq\lq shell\rq\rq\ and ii) a larger, bow-shaped shell at the southwestern region, called the \lq\lq wing\rq\rq\ \citep{LAN1982} (see Fig.\ref{fig:3}a). Due to its complex morphology, it has drawn the attention of the scientific community attempting to explain its overall shape. According to \cite{PIN1987}, the initial explosion of the VRO occurred in a region characterized by a density discontinuity. The part of the remnant that evolved into the denser region created the \lq\lq shell\rq\rq~component, while the rest diffused into a hotter and tenuous medium and shaped the \lq\lq wing\rq\rq~component. However, according to the recent results presented by \cite{ARI2019b}, there is no physical proof of an interaction of the remnant with the surrounding molecular clouds and they attributed the almost triangular shape of the \lq\lq wing\rq\rq\ to a Mach cone cavity which was created by a supersonically moving progenitor star and was filled out by the SN ejecta. Finally, \cite{CHIOT2019} modeled the observed morphology of VRO with 2D hydrodynamic simulations, suggesting that the remnant is currently interacting with the density wall of a wind bubble sculptured by the equatorial confined mass outflows of a supersonically moving progenitor star.

The interpretation of the remnant's morphology still remains an open issue. Our aim is, using the physical results presented below that deduced from the 3D MK model, to contribute to the clarification of unanswered questions with respect to this intriguingly complicated SNR.

\section{Observations}
\begin{figure*}
 \begin{center}
  \includegraphics[scale=0.35]{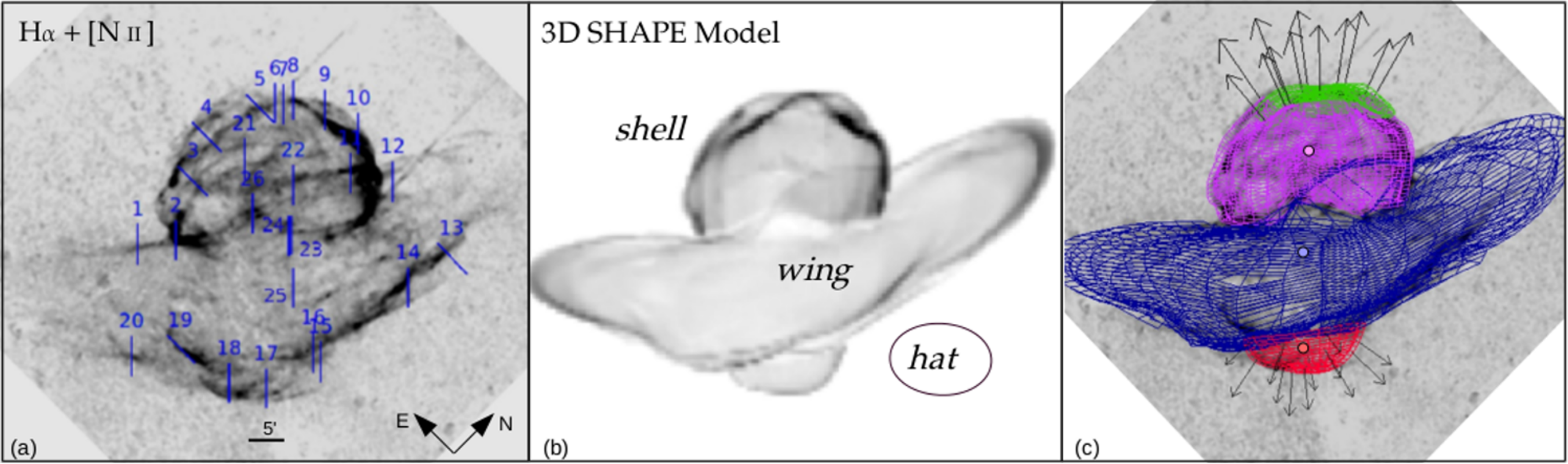}
  \caption{Fig.1a shows the VRO in \HNII ~emission lines. The blue labelled lines represent the slits' positions where high resolution spectra were obtained. In Fig.1b the 3D model of VRO is demonstrated, where its three components are labelled, and Fig.1c illustrates the 3D model in mesh-grid representation overlaid upon the \hnii\ image, without the slits' positions. The different colours correspond to the distinct components of the remnant with respect to their morphology and kinematics. The black arrows point to the direction of expansion of the green region and the \lq\lq hat\rq\rq ~component of VRO. The three coloured dots illustrate the geometrical centre of each component, which coincides with the centre of each component's velocity field too.}
     \label{fig:3} 
 \end{center}
  \end{figure*}

{\bf Wide-field and high-resolution imaging:}
Wide-field optical images covering the whole area of VRO (55 $\times$ 35 arcmin$^2$, with an image scale of 4 arcsec pixel$^{-1}$) were obtained with the $0.3$ m telescope at Skinakas Observatory (Greece) in 2000 and 2001 with the use of~\HNII, \OIII\ and \SII\ interference filters (\citealt{BOUMIS2012, BOUMIS2016}, 2020 in prep.). For the needs of the presented 3D model, the image in~\hnii\ filter was used due to the high brightness of the remnant in this emission line (see Fig.\ref{fig:3}a). Furthermore, high-resolution~\hnii\ images of selected areas of VRO have been obtained with the $2.3$ m Aristarchos telescope at Helmos Observatory (Greece) between 2011 and 2019 (Boumis et al. 2020 in prep.). Although these images were not digitized into the model constraints, they were used for the 3D visualization of VRO as they depict in great detail the filamentary structures of the remnant.\\
{\bf High-dispersion long-slit spectroscopy:} High-resolution long-slit echelle spectra were obtained in \HNII, \OIII\ and \SII\ between the years 2010 and 2019 (Boumis et al. 2020 in prep.) at the 2.1 m telescope in San Pedro Martir Observatory, Mexico, with the Manchester Echelle Spectrometer (MES-SPM; ~\citealt{MEA2003}). A 2048 $\times$ 2048 (13.5$\mu$m pixel size) CCD was used with a two times binning in both the spatial and spectral dimensions, resulting in a 0.35 arcsec pixel$^{-1}$ spatial scale. The slit length corresponds to 5.5 arcmin on the sky, while the slit width used was 300 $\mu$m (20 km s$^{-1}$ or 3.9 arcsec wide or 0.44 \AA). In total, 26 long-slit spectra were obtained with the purpose to cover the key areas of the remnant. From these, 21 spectra  were obtained with slit  Position Angle (P.A.) of 45$\degr$ with respect to the North and the rest 5 at a P.A. of 90$\degr$ (Fig.\ref{fig:3}a). The spectra were finally calibrated in heliocentric radial velocity ($V_{hel}$) to $\pm2.6$\,km\,s$^{-1}$ accuracy against spectra of a Th-Ar lamp. The data were analyzed in the standard way using the {\sc iraf} software package. From the spectra analysis, the kinematical information was obtained which was necessary for the needs of the 3D model.

\begin{figure*}
\begin{center}
\includegraphics[scale=0.28]{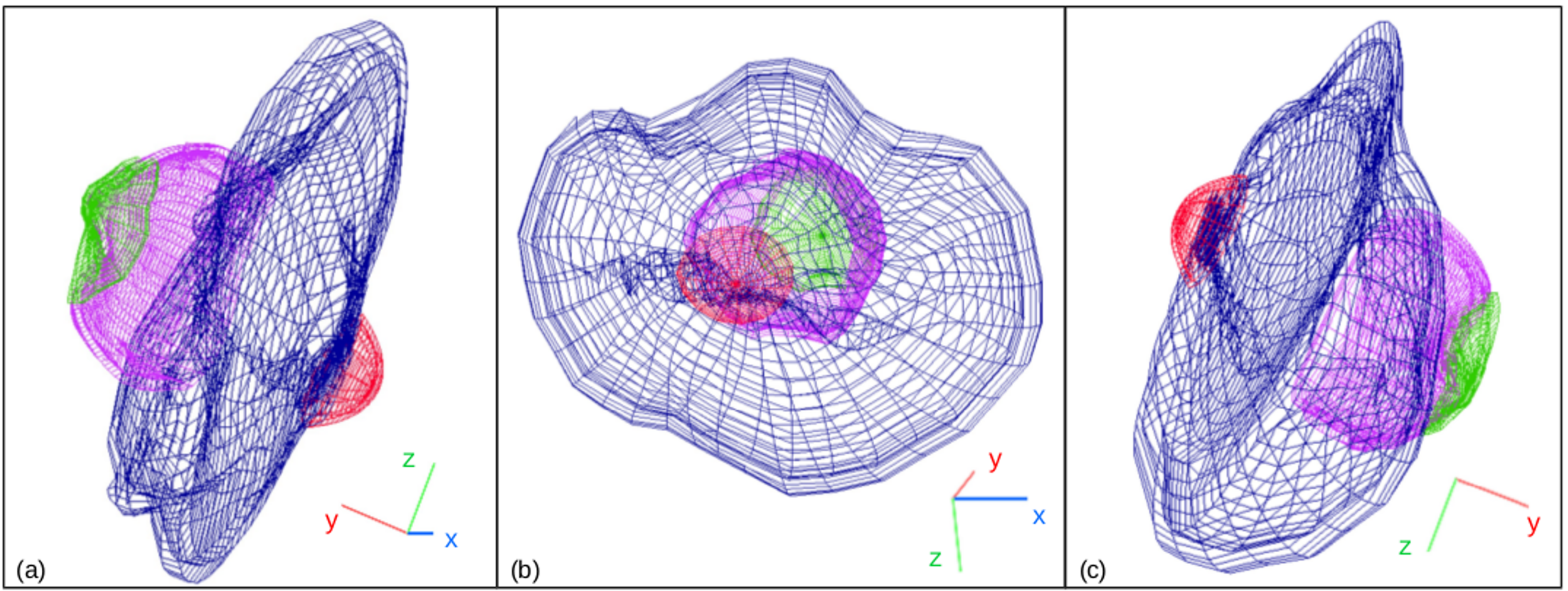}
\caption{3D model of VRO in mesh-grid representation, as seen from different angles rotated through x, y and z axis. For colour description see Fig.\ref{fig:3}. \href{https://drive.google.com/file/d/1_bqPMgrDej5G19hBWmZePVW5U_Refk9o/view?usp=sharing}{An animation of the 3D model can be found in the supporting materials of this paper}.}
\label{fig:1mesh} 
\end{center}
\end{figure*}

\begin{figure*}
   \begin{center}
    \includegraphics[scale=0.35]{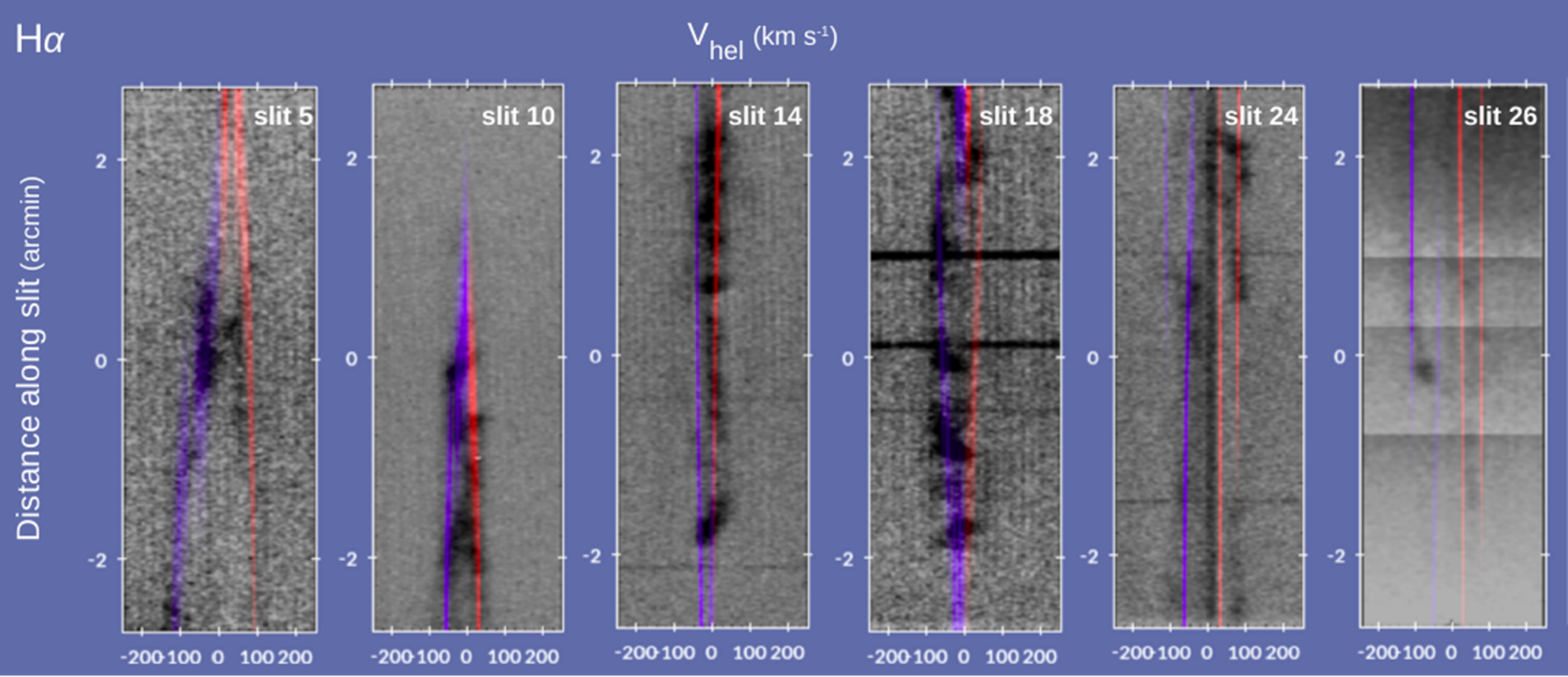}
    \caption{PV observational (in H$\alpha$ emission line) and synthetic diagrams of six different regions of VRO. For description see Section \ref{Results}.}
       \label{fig:pv} 
   \end{center}
    \end{figure*}

 \section{3D Morpho-Kinematical (MK) Modelling}
 \label{3D Modelling}
 
A 3D study of extended emission line sources like PNe, H~\rm{\sc ii} regions and SNRs is essential to provide new insights to their formation and evolution. Although PNe have been extensively studied via 3D MK modeling, there is no similar work for SNRs. The reasons for this are i) SNRs are usually very extended sources, which means that, apart from the imaging data, a great amount of spectroscopic data is also required for a full coverage of the remnant, ii) the asymmetries that SNRs usually display in their shape denote a great complexity in morphology and kinematics which is an additional factor of difficulty in the 3D visualization and iii) many SNRs are thin shells with low central emission along the line of sight, therefore most of the optical emission is tangent to the observer and hence radial velocities are close to zero. Below, we present the first 3D MK model for a SNR (VRO 42.05.01), using SHAPE and briefly outline the method.

For the reconstruction of the 3D structure of VRO,  H$\alpha$+[N {\sc ii}] images were used to constrain its 2D morphology projected on the plane of the sky. For the third dimension along the line of sight, the position-velocity (PV) diagrams of high-dispersion H$\alpha$ spectra from 26 regions were employed. We started by building the remnant from the \lq\lq wing\rq\rq ~component considering a spherical structure, which we gradually deformed applying a number of geometric tools in SHAPE (size, squeeze, bump). The other two components, \lq\lq shell\rq\rq ~and \lq\lq hat\rq\rq, were simulated as spherical shell structures with a finite thickness.
 
A key assumption for 3D MK modelling is the velocity field of a structure that allows us to constrain its size in the direction perpendicular to the plane of the sky. The homologous expansion law $\overrightarrow{V}$ = $B(\frac{\overrightarrow{r}}{Ro})$, where $r$ is the distance in arcmin of a given point from the centre of the field, $Ro$ is the distance in arcmin at which the velocity is equal to $B$ (km~s$^{-1}$), was used for all the components of VRO, with expansion centres for the \lq\lq shell\rq\rq ~and \lq\lq hat\rq\rq ~that are offset from that of the \lq\lq wing\rq\rq ~(see Fig.\ref{fig:3}c) and with different $B$ coefficients. The model provides us with a 3D-snapshot in time of the changing structure, while it is distance independent.

In case of PNe models, the centre of the velocity field coincides with the geometric centre of the nebula. However, in complex SNRs such as VRO with three different structures is more difficult to define the centre, given the fact that the position of the explosion is uncertain. Therefore in our model, we assumed that the expansion velocity field centre coincides with the geometrical center of each component. The expansion velocity laws of the  \lq\lq shell\rq\rq~($B$=115~km~s$^{-1}$, $Ro$=15.5 arcmin) and its upper part (green region in Fig.\ref{fig:3}c) ($B$=155~km~s$^{-1}$, $Ro$=15.5 arcmin), and of the \lq\lq hat\rq\rq ~($B$=90~km~s$^{-1}$, $Ro$=10 arcmin) structures were determined by matching the model with the observations from various slit positions. Due to the complex morphology of the \lq\lq wing\rq\rq, the velocity law was constrained using only the PV diagram of slit 25 (Fig.\ref{fig:3}a). Because of the position of slit 25, the angle between the expansion and radial velocities is small enough to minimize the effect of the inclination and provides a less uncertain velocity field ($B$=115~km~s$^{-1}$, $Ro$=39 arcmin). For the remaining slits, we consider the same velocity law and we deformed the shape and size of the structure until a satisfactory matching was obtained between the model and the observations.

Apart from the kinematics, the filamentary structures that VRO presents were also intriguing. As shown in Fig.\ref{fig:3}a, the remnant shows a filamentary structure, but especially in the \lq\lq shell\rq\rq\ ~there is network of filaments, crossing the entire surface. The most intense of them were reproduced as indentations on the surface of the component (see Fig.\ref{fig:3}b). \cite{PIN1987} had also characterized the\lq\lq shell\rq\rq ~as \lq\lq surface spherical in grand design but indented in detail\rq\rq. Furthermore, in the model we used the wing's outer edges as closed-ended, even though in Fig.\ref{fig:3}a it seems that they are open-ended. A higher contrast version of this image, shows that even they are fainter and not continuous, they are close-ended, but the possibility of having shock break-out features there cannot be ruled out.

\section{Results}
\label{Results}
Fig.\ref{fig:3}a illustrates the observational \hnii\ image of VRO along with the slits' positions, while in Fig.\ref{fig:3}b the 3D model of VRO is presented. Apart from the \lq\lq shell\rq\rq ~and the \lq\lq wing\rq\rq ~components, a third component has also been added in the southwestern region of the remnant, that we term it as the \lq\lq hat\rq\rq. The reason for representing this part as a separated structure was that, according to the observational data and the model, this lower part of the \lq\lq wing\rq\rq ~protrudes with respect to the rest \lq\lq wing\rq\rq, and also shows a different kinematic behaviour. In Fig.\ref{fig:3}c, the 3D visualization is illustrated but in a mesh representation, overlaid upon Fig.\ref{fig:3}a without the slits positions. The different colours correspond to the distinct components of the remnant, each one of which is characterized by its own morphology and velocity field. Fig.\ref{fig:1mesh} also presents the mesh representation of VRO, as seen from different directions.

We found that the whole remnant is tilted by approximately $\sim$3$\degr$-$5$\degr\ with respect to the plane of the sky. That means that the \lq\lq shell\rq\rq~goes inwards the page, while the \lq\lq hat\rq\rq ~component goes outwards from the page. Concerning the \lq\lq wing\rq\rq~component, the model showed that its northern part is bent with respect to its eastern counterpart, implying a possible interaction with a denser ambient medium at this part of the remnant (see also \citealt{ARI2019b}). In addition, a part of the \lq\lq wing\rq\rq~penetrates the central region of the \lq\lq shell\rq\rq ~at its front side (see Fig.\ref{fig:3}a near slit $26$), but also at its back side as well as showed by the model. Furthermore, the straight filament that crosses the shell in the positions of slits 11 and 22, is attributed - based on our model - to the northern back side of the \lq\lq wing\rq\rq ~component. The systemic velocity of the SNR was calculated to be $V_{\rm sys}$~=~-17$\pm 3~$\,km\,s$^{-1}$, lower than the value of $V_{\rm sys}$~=~$-34~$\,km\,s$^{-1}$ proposed by \cite{LAN1989}.

With regards to the expansion velocities of each component, the model showed that VRO is not characterized by a uniform velocity law. The \lq\lq shell\rq\rq ~appears to expand at a velocity of $V_{\rm exp}$~=~115$\pm 5~$\,km\,s$^{-1}$. However, the upper part of the \lq\lq shell\rq\rq ~(green region in Fig.\ref{fig:3}c) presents an expansion velocity towards north-east of $V_{\rm exp}$~=~155$\pm 15$ \,km\,s$^{-1}$, which is higher than that of the rest \lq\lq shell\rq\rq, and also corresponds to the the highest velocity of the remnant in total. The \lq\lq wing\rq\rq ~appears to have $V_{\rm exp}$~=~115$\pm 5~$\,km\,s$^{-1}$, same as the \lq\lq shell\rq\rq ~counterpart. On the opposite side of the remnant in the south-west region, the \lq\lq hat\rq\rq~was found to expand at a velocity of $V_{\rm exp}$~=~90$\pm 20~$\,km\,s$^{-1}$, which is lower than that of the \lq\lq wing\rq\rq~component. The black arrows in Fig.\ref{fig:3}c point to the direction of the radial expansion of the green region and the \lq\lq hat\rq\rq~of VRO. The velocities of  our \oiii\ spectra in these regions, i.e. slits 5 and 8 (V$\sim$120-130 \,km\,s$^{-1}$) and 18 (V$\sim$70-100 \,km\,s$^{-1}$), agree with those deduced from our model, which in turn are consistent with the velocities of theoretical shock models \citep{HART1987}.

Apart from the morphological resemblance of the reproduced model with the observational image, our guide in order to check the validity of our model was the overall agreement between the observational PV and the synthetic PV diagrams produced with SHAPE. In Fig.\ref{fig:pv}, six PV observational diagrams (black lines in H$\alpha$) from six characteristic regions of VRO are presented along with the synthetic coloured PVs reproduced with SHAPE. These spectra correspond to regions of the \lq\lq shell\rq\rq ~(slits $5$, $10$), the \lq\lq wing\rq\rq ~(slit $14$), the \lq\lq hat\rq\rq ~(slit $18$), and the contact regions between the \lq\lq shell\rq\rq~ and the \lq\lq wing\rq\rq~ (slits $24$ and $26$). The blue and red lines correspond to the blue-and red- shifted part of the remnant, respectively. At the points where the slit's position covers the regions of two components, see for example slits $24$ and $18$ in Fig.\ref{fig:3}a, the contribution in the synthetic spectra comes from both components. This is why there are two pairs  of blue-red lines in the synthetic PVs of these slits. Similarly, in the synthetic PV of slit $26$, both the \lq\lq wing\rq\rq~and the back side of the \lq\lq shell\rq\rq~contribute to the reproduced, synthetic PV diagram. The matching is quite sufficient, and it has been achieved for all the $26$ spectra obtained for VRO. The goal was the overall fitting between observational and synthetic spectra, neglecting at this point individual substructures (blobs etc.) that the spectra may illustrate. Therefore, the model results are consistent with the observations in both imaging and spectroscopic data.

 \section{Discussion}
 \label{Discussion}
 Due to its peculiar morphology, VRO has become the subject of investigation for many years, in an attempt for its morphology to be correctly interpreted. 

According to our model, VRO consists of three basic components: a \lq\lq shell\rq\rq, a \lq\lq wing\rq\rq ~and a \lq\lq hat\rq\rq. This distinction is on the basis of their morphology and kinematics. The first two structures were adopted from the already known literature and proved to have the same velocities range ($V_{\rm exp}$~=~115$\pm 5~$\,km\,s$^{-1}$), while the third structure was added in the model due to its different kinematics ($V_{\rm exp,hat}$~=90$\pm 20~$\,km\,s$^{-1}$) and the protrusion it presents with respect to the \lq\lq wing\rq\rq ~component. Finally, we found that, although the \lq\lq shell\rq\rq seems to be morphologically unified, its upper part (green region in Fig.\ref{fig:3}c) expands with a higher velocity of $V_{\rm exp}$~=~155$\pm 15~$\,km\,s$^{-1}$.

 Our 3D MK model showed that the remnant’s morphology displays a roughly axial symmetry in the azimuthal and polar dimension. This result advocates that VRO most likely was  shaped under an  axis or central symmetric  mechanism linked to the nature and evolution of the progenitor system.  From this perspective our results are aligned to the wind blown medium around the VRO remnant suggested by  \cite{CHIOT2019}. Within the framework of this model, the similar velocities that the \lq\lq shell\rq\rq ~and the \lq\lq wing\rq\rq ~display –despite their different shapes and sizes- can be attributed to the deceleration of the remnant caused by the collision of the SN blast wave with the density walls of the wind bubble. In the fast expanding upper part of the \lq\lq shell\rq\rq ~(green region) we may witness a shock breakout, where the blast wave penetrated the CSM density wall and is currently propagating in the lower density ambient medium. Finally, the \lq\lq hat\rq\rq ~component coincides with the region of the bow shaped CSM where the stagnation point is lying. The high circumstellar densities expected in the area of the stagnation point \citep[e.g][]{Chiotellis12} is aligned with the low velocities we gain from the \lq\lq hat\rq\rq ~component of the remnant.
  
An ISM density  discontinuity  suggested by \cite{PIN1987} could also be possible to explain the VRO properties as extracted by our 3D MK modeling. Within this model the \lq\lq wing\rq\rq~ had evolved in the low density region of the ISM and thus, it gained its extended size compared to the \lq\lq shell\rq\rq. Currently, one may say that the \lq\lq shell\rq\rq ~and the \lq\lq wing\rq\rq ~have swept up about the same mass and as a result they display similar expansion velocities. However, an extra ISM density gradient toward the NE and SW direction is required in order to explain the high and low velocity of the upper \lq\lq shell\rq\rq ~and the \lq\lq hat\rq\rq, respectively.

 \section{Conclusion}
 We present for the first time a 3D Morpho-Kinematic model of a SNR, VRO 42.05.01. The principal conclusions from this study are:
 \begin{enumerate}
     \item VRO can be represented by three basic distinct components, i.e. a \lq\lq shell\rq\rq, a \lq\lq wing\rq\rq, and a \lq\lq hat\rq\rq, each one of which presents specific morphological and kinematical characteristics. 
     \item The \lq\lq shell\rq\rq ~and the \lq\lq wing\rq\rq ~reveal similar expansion velocities of $V_{\rm exp}$~=~115$\pm 5~$\,km\,s$^{-1}$ while the \lq\lq hat\rq\rq ~is expanding with $V_{\rm exp}$~=~90$\pm 20~$\,km\,s$^{-1}$. Finally, the upper part of the \lq\lq shell\rq\rq ~displays the higher expansion velocity of the SNR equal to $V_{\rm exp}=$ 155$\pm 15~$\,km\,s$^{-1}$.
   \item The remnant has an inclination of $\sim$3$\degr$ - $5$\degr with respect to the plane of the sky and a systemic velocity of $V_{\rm sys}$~=~-17$\pm 3~$\,km\,s$^{-1}$.
     \item The northern part of the \lq\lq wing\rq\rq~component is tilted with respect to its eastern counterpart, due to a possible interaction with a denser ambient medium in this region of the SNR. 
     \item Our results are in line with the wind-bubble interaction model suggested by \cite{CHIOT2019}, however, a local ISM discontinuity in the vicinity of VRO suggested by \cite{PIN1987} cannot be excluded.
 \end{enumerate}

 \section*{Acknowledgements}
 
The authors would like to thank the referee for his/her thorough comments that improved the manuscript. S.D. and A.C. acknowledge the support of this work by the PROTEAS
II project (MIS 5002515), which is implemented under the
“Reinforcement of the Research and Innovation Infrastructure” action,
funded by the “Competitiveness, Entrepreneurship and Innovation”
operational programme (NSRF 2014-2020) and co-financed
by Greece and the European Union (European Regional Development
Fund). S.D acknowledges the Operational Programme “Human
Resources Development, Education and Lifelong Learning”
in the context of the project “Strengthening Human Resources Research
Potential via Doctorate Research” (MIS-5000432), implemented
by the State Scholarships Foundation (IKY) and co-financed
by Greece and the European Union (European Social Fund- ESF).
P.B and A.C. acknowledge the support of this work by the Operational
Program “Human Resources Development, Education and
Lifelong Learning 2014-2020” co-financed by Greece and the European
Union (European Social Fund-ESF) in the context of the
project “On the interaction of Type Ia Supernovae with Planetary
Nebulae” (MIS 5049922). W.S. was supported by UNAM DGAPA
PASPA. We would like to thank J. Dickel for informing us about the
origin of the name of VRO 42.05.01, from its detection at the Vermilion
River Observatory \citep{DICK1965}. This paper is based
on observations carried out at the OAN-SPM (México), Skinakas
Observatory (Crete, Greece) and Aristarchos telescope (Helmos,
Greece).

 \section*{Data Availability Statement}
 The data underlying this article will be shared on reasonable request to the corresponding author.




\bibliographystyle{mnras}
\bibliography{vro_shape.bib} 

\begin{thebibliography}{}
\makeatletter
\relax
\def\mn@urlcharsother{\let\do\@makeother \do\$\do\&\do\#\do\^\do\_\do\%\do\~}
\def\mn@doi{\begingroup\mn@urlcharsother \@ifnextchar [ {\mn@doi@}
  {\mn@doi@[]}}
\def\mn@doi@[#1]#2{\def\@tempa{#1}\ifx\@tempa\@empty \href
  {http://dx.doi.org/#2} {doi:#2}\else \href {http://dx.doi.org/#2} {#1}\fi
  \endgroup}
\def\mn@eprint#1#2{\mn@eprint@#1:#2::\@nil}
\def\mn@eprint@arXiv#1{\href {http://arxiv.org/abs/#1} {{\tt arXiv:#1}}}
\def\mn@eprint@dblp#1{\href {http://dblp.uni-trier.de/rec/bibtex/#1.xml}
  {dblp:#1}}
\def\mn@eprint@#1:#2:#3:#4\@nil{\def\@tempa {#1}\def\@tempb {#2}\def\@tempc
  {#3}\ifx \@tempc \@empty \let \@tempc \@tempb \let \@tempb \@tempa \fi \ifx
  \@tempb \@empty \def\@tempb {arXiv}\fi \@ifundefined
  {mn@eprint@\@tempb}{\@tempb:\@tempc}{\expandafter \expandafter \csname
  mn@eprint@\@tempb\endcsname \expandafter{\@tempc}}}

\bibitem[\protect\citeauthoryear{{Abell{\'a}n} et~al.,}{{Abell{\'a}n}
  et~al.}{2017}]{ABE2017}
{Abell{\'a}n} F.~J.,  et~al., 2017, \mn@doi [\apjl] {10.3847/2041-8213/aa784c},
  \href {https://ui.adsabs.harvard.edu/abs/2017ApJ...842L..24A} {842, L24}

\bibitem[\protect\citeauthoryear{{Akras} \& {L{\'o}pez}}{{Akras} \&
  {L{\'o}pez}}{2012}]{akras2012b}
{Akras} S.,  {L{\'o}pez} J.~A.,  2012, \mn@doi [\mnras]
  {10.1111/j.1365-2966.2012.21578.x}, \href
  {https://ui.adsabs.harvard.edu/abs/2012MNRAS.425.2197A} {425, 2197}

\bibitem[\protect\citeauthoryear{{Akras} \& {Steffen}}{{Akras} \&
  {Steffen}}{2012}]{akras2012a}
{Akras} S.,  {Steffen} W.,  2012, \mn@doi [\mnras]
  {10.1111/j.1365-2966.2012.20928.x}, \href
  {https://ui.adsabs.harvard.edu/abs/2012MNRAS.423..925A} {423, 925}

\bibitem[\protect\citeauthoryear{{Akras}, {Clyne}, {Boumis}, {Monteiro},
  {Gon{\c{c}}alves}, {Redman}  \& {Williams}}{{Akras} et~al.}{2016}]{akras2016}
{Akras} S.,  {Clyne} N.,  {Boumis} P.,  {Monteiro} H.,  {Gon{\c{c}}alves}
  D.~R.,  {Redman} M.~P.,   {Williams} S.,  2016, \mn@doi [\mnras]
  {10.1093/mnras/stw038}, \href
  {https://ui.adsabs.harvard.edu/abs/2016MNRAS.457.3409A} {457, 3409}

\bibitem[\protect\citeauthoryear{{Alarie}, {Bilodeau}  \& {Drissen}}{{Alarie}
  et~al.}{2014}]{ALA2014}
{Alarie} A.,  {Bilodeau} A.,   {Drissen} L.,  2014, \mn@doi [\mnras]
  {10.1093/mnras/stu774}, \href
  {https://ui.adsabs.harvard.edu/abs/2014MNRAS.441.2996A} {441, 2996}

\bibitem[\protect\citeauthoryear{{Arias} et~al.,}{{Arias}
  et~al.}{2019a}]{ARI2019a}
{Arias} M.,  et~al., 2019a, \mn@doi [\aap] {10.1051/0004-6361/201833865}, \href
  {https://ui.adsabs.harvard.edu/abs/2019A&A...622A...6A} {622, A6}

\bibitem[\protect\citeauthoryear{{Arias}, {Dom{\v{c}}ek}, {Zhou}  \&
  {Vink}}{{Arias} et~al.}{2019b}]{ARI2019b}
{Arias} M.,  {Dom{\v{c}}ek} V.,  {Zhou} P.,   {Vink} J.,  2019b, \mn@doi [\aap]
  {10.1051/0004-6361/201935528}, \href
  {https://ui.adsabs.harvard.edu/abs/2019A&A...627A..75A} {627, A75}

\bibitem[\protect\citeauthoryear{{Bolte}, {Sasaki}  \& {Breitschwerdt}}{{Bolte}
  et~al.}{2015}]{BOLT2015}
{Bolte} J.,  {Sasaki} M.,   {Breitschwerdt} D.,  2015, \mn@doi [\aap]
  {10.1051/0004-6361/201526000}, \href
  {https://ui.adsabs.harvard.edu/abs/2015A&A...582A..47B} {582, A47}

\bibitem[\protect\citeauthoryear{{Boumis}, {Alikakos}  \&
  {Mavromatakis}}{{Boumis} et~al.}{2012}]{BOUMIS2012}
{Boumis} P.,  {Alikakos} I.,   {Mavromatakis} F.,  2012, in {Papadakis} I.,
  {Anastasiadis} A.,  eds, 10th Hellenic Astronomical Conference. p.~27

\bibitem[\protect\citeauthoryear{{Boumis}, {Akras}, {Leonidaki}, {Chiotellis},
  M., {Alikakos}, N.  \& {Mavromatakis}}{{Boumis} et~al.}{2016}]{BOUMIS2016}
{Boumis} P.,  {Akras} S.,  {Leonidaki} I.,  {Chiotellis} A.,  M. K.,
  {Alikakos} I.,  N. N.,   {Mavromatakis} F.,  2016, in {Boumis} P.,  {Raymond}
  J.,  eds, Supernova Remnants: An Odyssey in Space after Stellar Death
  Conference. p.~15

\bibitem[\protect\citeauthoryear{{Chiotellis}, {Schure}  \&
  {Vink}}{{Chiotellis} et~al.}{2012}]{Chiotellis12}
{Chiotellis} A.,  {Schure} K.~M.,   {Vink} J.,  2012, \mn@doi [\aap]
  {10.1051/0004-6361/201014754}, \href
  {http://adsabs.harvard.edu/abs/2012A%26A...537A.139C} {537, A139}

\bibitem[\protect\citeauthoryear{{Chiotellis}, {Boumis}, {Derlopa}  \&
  {Steffen}}{{Chiotellis} et~al.}{2019}]{CHIOT2019}
{Chiotellis} A.,  {Boumis} P.,  {Derlopa} S.,   {Steffen} W.,  2019, arXiv
  e-prints, \href {https://ui.adsabs.harvard.edu/abs/2019arXiv190908947C} {p.
  arXiv:1909.08947}

\bibitem[\protect\citeauthoryear{{Clyne}, {Akras}, {Steffen}, {Redman},
  {Gon{\c{c}}alves}  \& {Harvey}}{{Clyne} et~al.}{2015}]{CLY2015}
{Clyne} N.,  {Akras} S.,  {Steffen} W.,  {Redman} M.~P.,  {Gon{\c{c}}alves}
  D.~R.,   {Harvey} E.,  2015, \mn@doi [\aap] {10.1051/0004-6361/201526585},
  \href {https://ui.adsabs.harvard.edu/abs/2015A&A...582A..60C} {582, A60}

\bibitem[\protect\citeauthoryear{{DeLaney} et~al.,}{{DeLaney}
  et~al.}{2010}]{DEL2010}
{DeLaney} T.,  et~al., 2010, \mn@doi [\apj] {10.1088/0004-637X/725/2/2038},
  \href {https://ui.adsabs.harvard.edu/abs/2010ApJ...725.2038D} {725, 2038}

\bibitem[\protect\citeauthoryear{{Derlopa}, {Akras}, {Boumis}  \&
  {Steffen}}{{Derlopa} et~al.}{2019}]{DER2019}
{Derlopa} S.,  {Akras} S.,  {Boumis} P.,   {Steffen} W.,  2019, \mn@doi
  [\mnras] {10.1093/mnras/stz193}, \href
  {https://ui.adsabs.harvard.edu/abs/2019MNRAS.484.3746D} {484, 3746}

\bibitem[\protect\citeauthoryear{{Dickel}, {McGuire}  \& {Yang}}{{Dickel}
  et~al.}{1965}]{DICK1965}
{Dickel} J.~R.,  {McGuire} J.~P.,   {Yang} K.~S.,  1965, \mn@doi [\apj]
  {10.1086/148346}, \href
  {https://ui.adsabs.harvard.edu/abs/1965ApJ...142..798D} {142, 798}

\bibitem[\protect\citeauthoryear{{Fang}, {Zhang}, {Kwok}, {Hsia}, {Chau},
  {Ramos-Larios}  \& {Guerrero}}{{Fang} et~al.}{2018}]{FANG2018}
{Fang} X.,  {Zhang} Y.,  {Kwok} S.,  {Hsia} C.-H.,  {Chau} W.,  {Ramos-Larios}
  G.,   {Guerrero} M.~A.,  2018, \mn@doi [\apj] {10.3847/1538-4357/aac01e},
  \href {https://ui.adsabs.harvard.edu/abs/2018ApJ...859...92F} {859, 92}

\bibitem[\protect\citeauthoryear{{Filippenko}}{{Filippenko}}{1997}]{FIL1997}
{Filippenko} A.~V.,  1997, \mn@doi [\araa] {10.1146/annurev.astro.35.1.309},
  \href {https://ui.adsabs.harvard.edu/abs/1997ARA&A..35..309F} {35, 309}

\bibitem[\protect\citeauthoryear{{G{\'o}mez-Gordillo}, {Akras},
  {Gon{\c{c}}alves}  \& {Steffen}}{{G{\'o}mez-Gordillo}
  et~al.}{2020}]{Gordillo2020}
{G{\'o}mez-Gordillo} S.,  {Akras} S.,  {Gon{\c{c}}alves} D.~R.,   {Steffen} W.,
   2020, \mn@doi [\mnras] {10.1093/mnras/staa060}, \href
  {https://ui.adsabs.harvard.edu/abs/2020MNRAS.492.4097G} {492, 4097}

\bibitem[\protect\citeauthoryear{{Hartigan}, {Raymond}  \&
  {Hartmann}}{{Hartigan} et~al.}{1987}]{HART1987}
{Hartigan} P.,  {Raymond} J.,   {Hartmann} L.,  1987, \mn@doi [\apj]
  {10.1086/165204}, \href
  {https://ui.adsabs.harvard.edu/abs/1987ApJ...316..323H} {316, 323}

\bibitem[\protect\citeauthoryear{{Landecker}, {Pineault}, {Routledge}  \&
  {Vaneldik}}{{Landecker} et~al.}{1982}]{LAN1982}
{Landecker} T.~L.,  {Pineault} S.,  {Routledge} D.,   {Vaneldik} J.~F.,  1982,
  \mn@doi [\apjl] {10.1086/183885}, \href
  {https://ui.adsabs.harvard.edu/abs/1982ApJ...261L..41L} {261, L41}

\bibitem[\protect\citeauthoryear{{Landecker}, {Pineault}, {Routledge}  \&
  {Vaneldik}}{{Landecker} et~al.}{1989}]{LAN1989}
{Landecker} T.~L.,  {Pineault} S.,  {Routledge} D.,   {Vaneldik} J.~F.,  1989,
  \mn@doi [\mnras] {10.1093/mnras/237.1.277}, \href
  {https://ui.adsabs.harvard.edu/abs/1989MNRAS.237..277L} {237, 277}

\bibitem[\protect\citeauthoryear{{McKee}}{{McKee}}{1988}]{McKee1988}
{McKee} C.~F.,  1988, in {Roger} R.~S.,  {Landecker} T.~L.,  eds, IAU Colloq.
  101: Supernova Remnants and the Interstellar Medium. p.~205

\bibitem[\protect\citeauthoryear{{Meaburn}, {L{\'o}pez}, {Guti{\'e}rrez},
  {Quir{\'o}z}, {Murillo}, {Vald{\'e}z}  \& {Pedrayez}}{{Meaburn}
  et~al.}{2003}]{MEA2003}
{Meaburn} J.,  {L{\'o}pez} J.~A.,  {Guti{\'e}rrez} L.,  {Quir{\'o}z} F.,
  {Murillo} J.~M.,  {Vald{\'e}z} J.,   {Pedrayez} M.,  2003, \rmxaa, \href
  {http://adsabs.harvard.edu/abs/2003RMxAA..39..185M} {39, 185}

\bibitem[\protect\citeauthoryear{{Milisavljevic} \& {Fesen}}{{Milisavljevic} \&
  {Fesen}}{2013}]{MIL2013}
{Milisavljevic} D.,  {Fesen} R.~A.,  2013, \mn@doi [\apj]
  {10.1088/0004-637X/772/2/134}, \href
  {https://ui.adsabs.harvard.edu/abs/2013ApJ...772..134M} {772, 134}

\bibitem[\protect\citeauthoryear{{Orlando} et~al.,}{{Orlando}
  et~al.}{2019}]{ORL2019}
{Orlando} S.,  et~al., 2019, \mn@doi [\aap] {10.1051/0004-6361/201834487},
  \href {https://ui.adsabs.harvard.edu/abs/2019AA...622A..73O} {622, 73}

\bibitem[\protect\citeauthoryear{{Pineault}, {Landecker}  \&
  {Routledge}}{{Pineault} et~al.}{1987}]{PIN1987}
{Pineault} S.,  {Landecker} T.~L.,   {Routledge} D.,  1987, \mn@doi [\apj]
  {10.1086/165161}, \href
  {https://ui.adsabs.harvard.edu/abs/1987ApJ...315..580P} {315, 580}

\bibitem[\protect\citeauthoryear{{Potter}, {Staveley-Smith}, {Reville}, {Ng},
  {Bicknell}, {Sutherland}  \& {Wagner}}{{Potter} et~al.}{2014}]{POT2014}
{Potter} T.~M.,  {Staveley-Smith} L.,  {Reville} B.,  {Ng} C.~Y.,  {Bicknell}
  G.~V.,  {Sutherland} R.~S.,   {Wagner} A.~Y.,  2014, \mn@doi [\apj]
  {10.1088/0004-637X/794/2/174}, \href
  {https://ui.adsabs.harvard.edu/abs/2014ApJ...794..174P} {794, 174}

\bibitem[\protect\citeauthoryear{{Steffen} \& {Koning}}{{Steffen} \&
  {Koning}}{2017}]{STEFF2017}
{Steffen} W.,  {Koning} N.,  2017, \mn@doi [Astronomy and Computing]
  {10.1016/j.ascom.2017.06.002}, \href
  {https://ui.adsabs.harvard.edu/abs/2017A&C....20...87S} {20, 87}

\bibitem[\protect\citeauthoryear{{Toledo-Roy}, {Vel{\'a}zquez}, {Esquivel}  \&
  {Giacani}}{{Toledo-Roy} et~al.}{2014}]{TOL2014G352}
{Toledo-Roy} J.~C.,  {Vel{\'a}zquez} P.~F.,  {Esquivel} A.,   {Giacani} E.,
  2014, \mn@doi [\mnras] {10.1093/mnras/stt1955}, \href
  {https://ui.adsabs.harvard.edu/abs/2014MNRAS.437..898T} {437, 898}

\bibitem[\protect\citeauthoryear{{Williams} et~al.,}{{Williams}
  et~al.}{2017}]{WIL2017}
{Williams} B.~J.,  et~al., 2017, \mn@doi [\apj] {10.3847/1538-4357/aa7384},
  \href {https://ui.adsabs.harvard.edu/abs/2017ApJ...842...28W} {842, 28}

\makeatother
\end{thebibliography}





\bsp	
\label{lastpage}
\end{document}